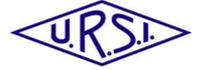

# An Alternative Method to Identify the Susceptibility Threshold Level of Device under Test in a Reverberation Chamber

Qian Xu*[(1)], Kai Chen[(1)(2)], Xueqi Shen[(2)], Lei Xing[(1)], Yi Huang[(3)], and Tian Hong Loh[(4)]
(1) Nanjing University of Aeronautics and Astronautics University, Nanjing, China; e-mail: emxu@foxmail.com
(2) Nanjing Rongce Testing Technology Ltd, Nanjing, China; e-mail: kay@emcdir.com; george@emcdir.com
(3) University of Liverpool, Liverpool, UK; e-mail: yi.huang@liverpool.ac.uk
(4) Electromagnetic & Electrochemical Technologies Department, Electromagnetic Technologies Group, National Physical Laboratory, Teddington, UK; e-mail: tian.loh@npl.co.uk

## Abstract

By counting the number of pass/fail occurrences of a DUT (Device under Test) in the stirring process in a reverberation chamber (RC), the threshold electric field (E-field) level can be well estimated without tuning the input power and repeating the whole testing many times. The Monte-Carlo method is used to verify the results. Estimated values and uncertainties are given for Rayleigh distributed fields and for Rice distributed fields with different *K*-factors.

## 1 Introduction

Conventionally, to find the susceptibility threshold electric field (E-field) level of device under test (DUT) in an anechoic chamber (AC) or a reverberation chamber (RC), one needs to tune the input power until the pass/fail can be well-identified. This process is relatively easy for testing in an anechoic chamber, as the incident E-field can be calculated or measured quickly in an AC.

However, for testing in an RC, E-fields at different positions can be statistically independent although they share the same probability density function (PDF), and the maximum E-field is described as a PDF. The maximum E-field in RCs can be statistically well described and has been applied to radiated susceptibility (RS) testing for many years [1]-[10]. It is easy to give a pass/fail conclusion for a given maximum E-field PDF in an RC. However, if one needs to find the threshold E-field level of a DUT in an RC, it can be very time-consuming. Since measurements in RC require the rotation of stirrers, applying the conventional tuning-and-repeating process in an RC can be counter-productive in process control for DUT evaluation. Fast methods have been proposed in [5]-[9] by taking advantage of statistical properties of RCs, counting the number of passes or fails is enough to find the threshold level. In this paper, we use a Monte-Carlo method to identify the unbiased estimator and the uncertainty of the estimated results, which provides a general approach for different independent sample numbers.

## 2 Theory and Simulations

In an RC the magnitude of E-field has a Rayleigh distribution with PDF given by

$$p(x) = \frac{x}{\sigma^2} e^{-\frac{x^2}{2\sigma^2}}, \quad x = |E_x| \qquad (1)$$

where $|E_x|$ is the magnitude of the E-field in *x*-, *y*- or *z*-polarization. The mean value of $|E_x|$ is $\sigma\sqrt{\pi/2}$ and the standard deviation is $\sigma\sqrt{2-\pi/2}$. The expected value of $|E_x|$ can be obtained from the net input power in an RC or the received power of an antenna in an RC [1]. Let $\sigma = \sqrt{2/\pi}$, the rectangular E-field is normalized to the mean value (1 V/m). Conventionally, from the extreme distributions, the expected value of the maximum E-field ($\langle \lceil |E_x| \rceil_N \rangle$) can be estimated as [1], [2]

$$\langle \lceil |E_x| \rceil_N \rangle \approx \sqrt{\frac{4}{\pi}\left[0.5772 + \ln(N+1) - \frac{1}{2(N+1)}\right]} \qquad (2)$$

where *N* is the number of independent stirrer positions, the probability of the confidence interval can be estimated using the cumulative distribution function (CDF) function of the maximum E-field [1], [2]

$$F_{\text{MaxEx}}(x) = \left[1 - \exp\left(\frac{-x^2}{2\sigma^2}\right)\right]^N \text{ where } \sigma = \sqrt{2/\pi} \qquad (3)$$

From (2) and (3), it can be found that although the confidence interval of maximum E-field can be quantified, when a DUT fails to pass the RS testing, the estimation of threshold of the DUT is not straightforward. One needs to repeat the measurement with different maximum E-field, and *N* independent samples are required in each measurement. This process can be very time-consuming. Although the confidence interval of the maximum E-field can be quantified, the confidence interval of the threshold level of a DUT is not easy to estimate. Fast methods which use the number of pass/fail results in the RS testing have been proposed [5]-[9]. From (1), the CDF of $|E_x|$ can be obtained as

$$F(x) = \int_0^x \frac{t}{\sigma^2} e^{-\frac{t^2}{\sigma^2}} dt \qquad (4)$$

Plots of (1) and (4) for $\sigma = \sqrt{2/\pi}$ are given in Figure 1. Suppose the threshold value of the DUT is $E_{\text{thr}}$ (normalized to the mean value of the rectangular E-field $\langle |E_x| \rangle$) as shown in Figure 1. Note that this threshold value in RC is different from the magnitude of the incident wave in AC. The conversion between them involves the directivity and the efficiency of the receiving antenna of the DUT [3]. This threshold value can be understood as the mean value of the E-field magnitude in a rich multipath environment with Rayleigh distribution. If the pass/fail status of the DUT can be recorded in all stirrer positions, the CDF value at $E_{\text{thr}}$ can be estimated as

$$F(E_{\text{thr}}) = \lim_{N \to \infty} N_{\text{low}}/N \qquad (5)$$

where $N_{\text{low}}$ is the number of pass results for the DUT (the sample E-field is lower than the threshold E-field), $N$ is the number of independent stirrer positions. It is assumed that the fail status of the DUT can be recovered when the external E-field is removed, and no permanent damage will occur in the RS testing [10]. Otherwise, one should start from very low input power and increase it slowly and carefully by following the conventional procedure. Intuitively, as long as $N_{\text{low}}/N$ is obtained, $E_{\text{thr}}$ can be solved from (5) directly. For finite $N$ values, (5) could be biased. A correction parameter $e(N, E_{\text{est}})$ would be necessary, and an unbiased estimation can be written as

$$E_{\text{thr}} = E_{\text{est}} e(N, E_{\text{est}}) \qquad (6)$$

where $E_{\text{est}} = \langle F^{-1}(N_{\text{low}}/N) \rangle$ represents the expected value of the estimated E-field. Obviously, when $N \to \infty$, $E_{\text{thr}} = E_{\text{est}}$. It can be observed from (6) that once $N_{\text{low}}$, $N$ and $e(N, E_{\text{est}})$ are known, $E_{\text{thr}}$ can be solved easily. Since $N_{\text{low}}$ and $N$ are obtained from measurements, one needs to find $e(N, E_{\text{est}})$ from theory or simulations. In this paper, Monte-Carlo simulations have been performed to evaluate $e(N, E_{\text{est}})$. The Monte-Carlo method is a general approach for solving models without closed-form expressions. The simulation procedure is illustrated in Figure 2, the step-by-step procedure is given as follows:

1) For a given $N$, generate $N$ samples with Rayleigh distributions randomly;
2) Give a threshold value $E_{\text{thr}}$, by counting the samples which are lower than $E_{\text{thr}}$ (the number of pass results), $N_{\text{low}}$ can be obtained. Counting the sample number $N_{\text{high}}$ which gives the number of fail results is equivalent since $N_{\text{low}} = N - N_{\text{high}}$.
3) Solve for the estimated threshold value $E_{\text{est1}} = F^{-1}(N_{\text{low}}/N)$;
4) Repeat 1) - 3) for the Monte-Carlo simulations $M = 10^5$ times, one obtains $E_{\text{est1}}, E_{\text{est2}}, \dots, E_{\text{estM}}$, find the mean ($E_{\text{est}} = \langle F^{-1}(N_{\text{low}}/N) \rangle$) and the standard deviations of the estimated threshold E-field samples.
5) Repeat 1) – 4) for different $N$ and $E_{\text{thr}}$ to find $e(N, E_{\text{est}})$ and the dependency of standard deviations.

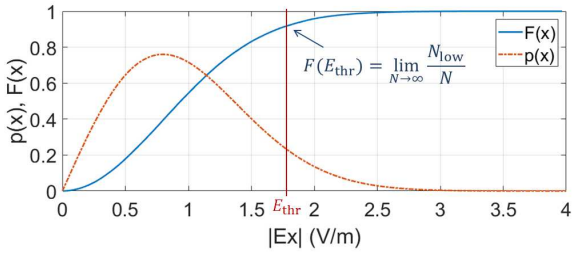

**Figure 1.** PDF and CDF plots of Rayleigh distribution with the expected value of 1V/m.

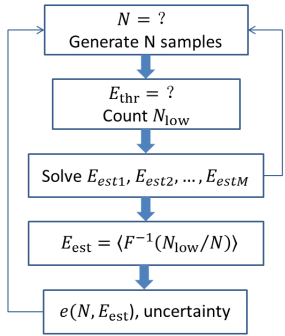

**Figure 2.** The flow chart of the Monte-Carlo simulation.

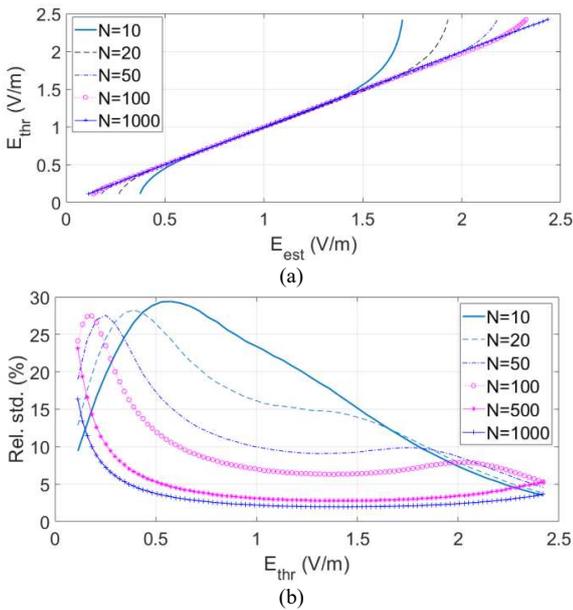

**Figure 3.** (a) The relationship between $E_{\text{est}}$ and $E_{\text{thr}}$, (b) the uncertainty of the final $E_{\text{thr}}$.

After performing Monte-Carlo simulations, the difference between the estimated threshold value $E_{\text{est}}$ and the given threshold value $E_{\text{thr}}$ can be identified. The simulated results are presented in Figure 3. The ratio of $E_{\text{thr}}$ and $E_{\text{est}}$ is illustrated in Figure 3(a), and the relative standard deviation (uncertainty) is given in Figure 3(b). As expected, when $N \to \infty$, $E_{\text{thr}} = E_{\text{est}}$.

In practical engineering, we use an example to demonstrate the use of Figure 3: suppose an RC has been calibrated with $\langle |E_x| \rangle = 50$ V/m, 10 stirrer positions were used in measurements ($N = 10$), in the RS testing, $N_{\text{low}} = 9$ stirrer positions were found to give pass results and 1 stirrer position was found to give fail results. Thus the estimated E-field threshold (biased value) can be obtained from $E_{\text{est}} = F^{-1}(N_{\text{low}}/N) = F^{-1}(0.9) \approx 1.71\,\text{V/m}$, by applying the correcting parameter in Figure 3(a), the threshold E-field (unbiased value) can be estimated as $E_{\text{thr}} \approx 2.4\,\text{V/m}$. Note that $E_{\text{thr}}$ is the normalized E-field, the unbiased threshold is finally estimated as $\langle|E_x|\rangle E_{\text{thr}} \approx 50 \times 2.4 = 120\,\text{V/m}$, the relative standard deviation (uncertainty) can be estimated as 3.5% from Figure 3(b).

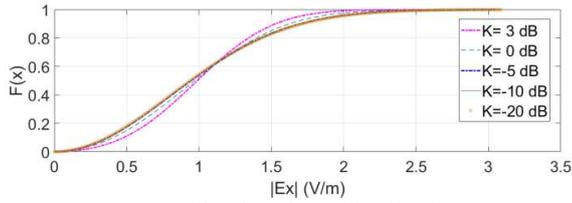

**Figure 4.** Normalized Rice distribution CDFs for different *K*-factors, the expected values are 1 V/m.

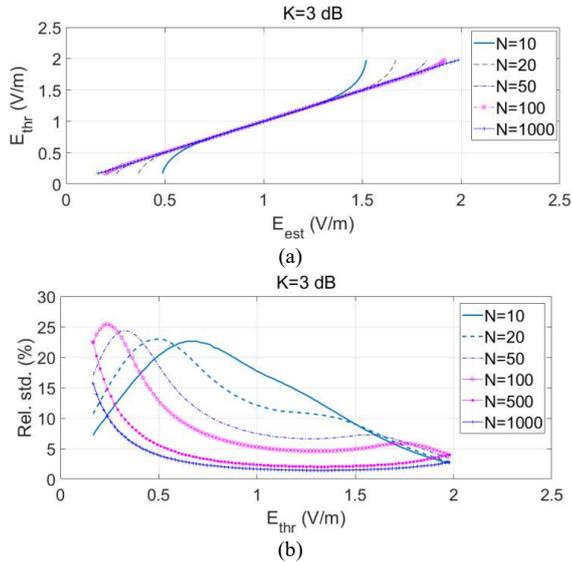

**Figure 5**. (a) The relationship between $E_{\text{est}}$ and $E_{\text{thr}}$, (b) The uncertainty of the final $E_{\text{thr}}$ with *K*=3 dB.

When the RC is not ideally stirred, or there is unstirred part (line of sight) between Tx and Rx antenna, a Rice distribution is assumed for $|E_x|$. The same process can be repeated for the Rice distribution with different *K*-factors. The normalized CDFs are illustrated in Figure 4 with the expected value of 1 V/m. Figure 5(a) – Figure 5(f) show the simulated $E_{\text{est}}$ and the uncertainties for different *K*-factors. It can be observed that when *K*-factor is lower than -5 dB, the results are very close to that of the Rayleigh distributions. The employed CDF range in simulations is 1% - 99%. In the Monte-Carlo simulations, the cases in which all E-field samples are lower or higher than the threshold have been excluded. Obviously, when all the samples are lower or higher than the threshold, one needs to tune the input power or increase *N* to find the threshold level. The probability that all the *N* samples are lower or higher than the given threshold are simulated and illustrated in Figure 6(a) and Figure 6(b), respectively. It is noted that it is possible that all the measured samples could be lower than the threshold. This effect is not easy to discover as the threshold level is unknown and no abnormality can be observed in practice. As expected, with the increase of *N*, the probabilities in Figure 6 are reduced.

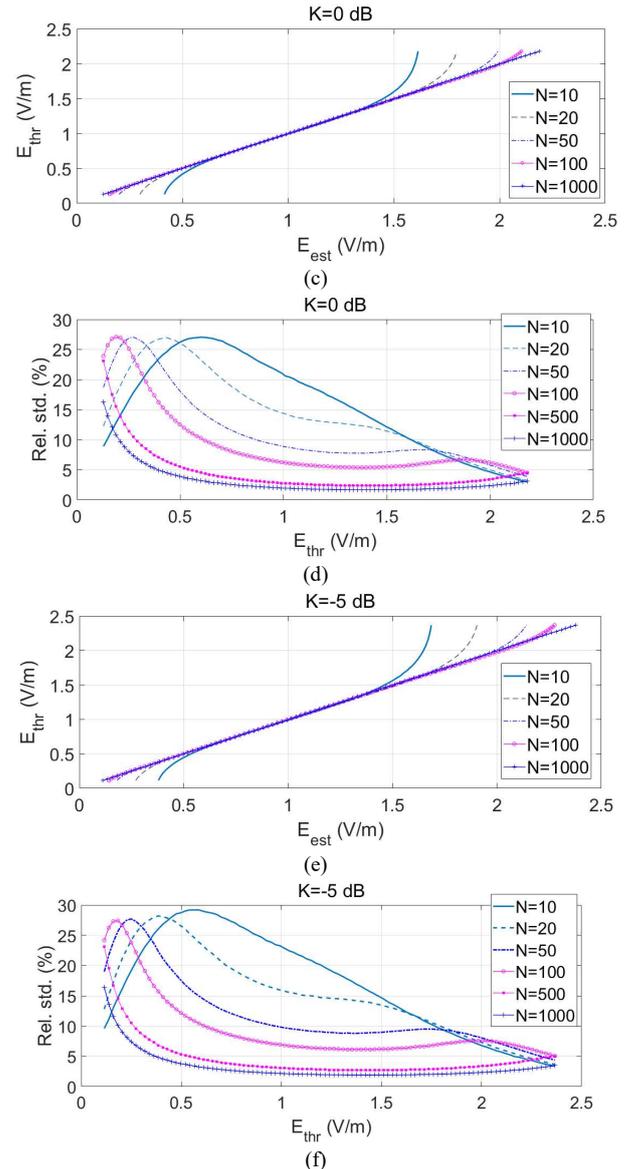

**Figure 5**. (c) The relationship between $E_{\text{est}}$ and $E_{\text{thr}}$, (d) the uncertainty of the final $E_{\text{thr}}$ with *K*=0 dB; (e) The relationship between $E_{\text{est}}$ and $E_{\text{thr}}$, (f) the uncertainty of the final $E_{\text{thr}}$ with *K*=-5 dB.

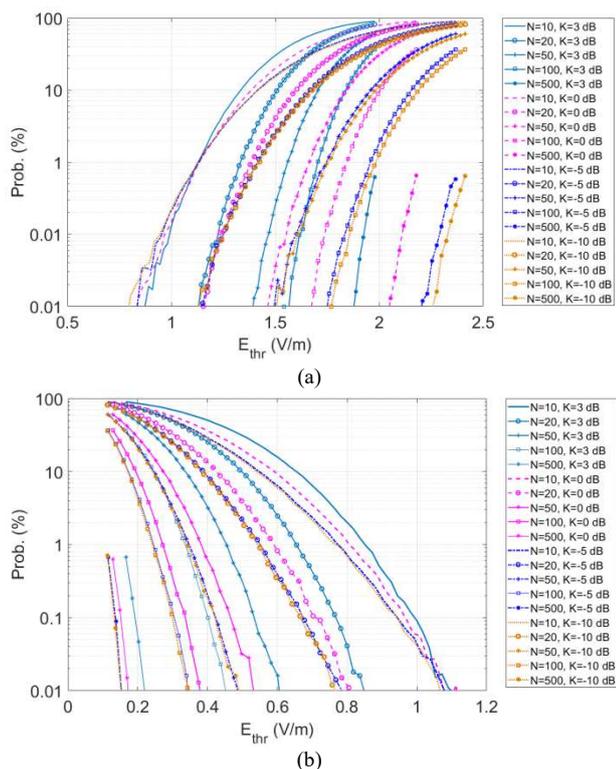

**Figure 6**. (a) The probability that all the *N* samples are lower than the given normalized threshold $E_{\text{thr}}$, (b) the probability that all the *N* samples are higher than the given normalized threshold $E_{\text{thr}}$.

## 3 Conclusions

The Monte-Carlo method has been used to quantify the unbiased estimator and the uncertainty of the estimated threshold level for different independent sample number. As long as the fail/pass status of the DUT is counting in the stirring process, the threshold level can be estimated with a given uncertainty. The results have also been generalized to Rice distributions with different *K*-factors.

## Acknowledgements

This work was supported in part by the Fundamental Research Funds for the Central Universities, Grant Number: NS2021029, Nanjing Rongce Testing Technology Ltd, 2021–2025 National Measurement System Programme of the UK Government's Department for Business, Energy and Industrial Strategy, Science Theme Reference EMT22 and EMT23, the EURAMET European Partnership on Metrology (EPM), under 21NRM03 Metrology for Emerging Wireless Standards (MEWS) project (The project 21NRM03 MEWS has received funding from the EPM, co-financed from the European Union's Horizon Europe Research and Innovation Programme and by the Participating States), the National Defense Basic Scientific Research Program of China under Grant JCKYS2021DC05 and the Fund of Prospective Layout of Scientific Research for NUAA.